\documentclass[twocolumn,showpacs,preprintnumbers,superscriptaddress,amsmath,amssymb]{revtex4}
\usepackage{amsfonts,amssymb,amscd,amsmath}
\usepackage{graphicx}

\DeclareMathOperator{\Ln}{Ln}

\newcommand{\slashit}[1]{#1 \kern-.45em\slash}

\newcommand{\slashP}{P \kern-.65em\slash }
\begin{document}
\title{Direct photons and dileptons via color dipoles}
%\author{To be confirmed}
\author{ B. Z. Kopeliovich}
\affiliation{Departamento de F\'\i sica y Centro de Estudios Subat\'omicos, Universidad
T\'ecnica Federico Santa Mar\'\i a, Casilla 110-V, Valpara\'\i so,
Chile}
\affiliation{Joint Institute for Nuclear Research, Dubna, Russia}
\author{A. H. Rezaeian}
\affiliation{Departamento de F\'\i sica y Centro de Estudios Subat\'omicos, Universidad
T\'ecnica Federico Santa Mar\'\i a, Casilla 110-V, Valpara\'\i so,
Chile}
\author{H. J. Pirner}
\affiliation{Institute for Theoretical Physics, University of Heidelberg,
Philosophenweg 19, D-69120 Heidelberg, Germany}
\author{Iv\'an Schmidt}
\affiliation{Departamento de F\'\i sica y Centro de Estudios Subat\'omicos, Universidad
T\'ecnica Federico Santa Mar\'\i a, Casilla 110-V, Valpara\'\i so,
Chile}
%\email{pir@tphys.uni-heidelberg.de}

\date{\today}
\begin{abstract}
Drell-Yan dilepton pair production and
inclusive direct photon production can be described within a unified
framework in the color dipole approach.  The inclusion of
non-perturbative primordial transverse momenta and DGLAP evolution
is studied. We successfully describe data for dilepton spectra from
$800$-GeV $pp$ collisions, inclusive direct
photon spectra for $pp$ collisions at RHIC energies $\sqrt{s}=200$
GeV, and for $p\bar{p}$ collisions at Tevatron energies $\sqrt{s}=1.8$
TeV, in a formalism that is free from any extra parameters.

\end{abstract}
%\keywords{given latter}
\maketitle
\date{\today}
\section{Introduction}
 Massive lepton pair production and inclusive direct photon production in hadronic
collisions have historically provided an important tool to gain
access to parton distributions in hadrons. Moreover, direct photons,
i.e. photons not from hadronic decay, can be also a powerful probe
of the initial state of matter created in heavy ion collisions,
since they interact with the medium only electromagnetically and
therefore provide a baseline for the interpretation of jet-quenching
models.

%Generally, the measured cross section
%lie above the theoretical predictions at low transverse momentum.
%One of problem
%in perturbative calculations of prompt-photon production is
%uncertainties associated with the renormalization, factorization and
%fragmentation schemes.
%In this paper, we investigate the inclusive
%prompt-photon and lepton pairs production in the color dipole approach
%which is free from the above problem.

In the parton model, the Feynman diagrams for partonic subprocesses
that are present in Drell-Yan (DY) lepton pair production and in
inclusive direct photon production are different, and the connection
between both production mechanisms within a unique approximation
scheme is not obvious. Since in the target rest frame the DY process
looks like bremsstrahlung of a virtual photon decaying into a lepton
pair, we will show that the color dipole formalism defined in this
frame is well suited to describe both production processes in a
unified framework free of parameters. As an illustrative example, we
study dilepton spectra in $800$-GeV $pp$ collisions from the E866
experiment \cite{p2}, inclusive direct-photon spectra in $pp$ at
$\sqrt{s}=200$ GeV from the PHENIX experiment \cite{p3}, and
$p\bar{p}$ collisions at $\sqrt{s}=1.8$ TeV from the CDF experiment
\cite{p4}.

There have been already some attempts to describe the DY transverse
momentum distribution in the color dipole approach \cite{p5}, but
unfortunately the experimental data that was used for comparison is
not fully kinematically in the range of validity of the model. Here we
confront the dipole approach with experimental data that is in a
region where the model is supposed to be at work.  Furthermore, we
will also study the inclusion of both non-perturbative primordial
transverse momenta and DGLAP evolution.

Despite many years of intense studies, a satisfactory description of
all existing inclusive direct photon production data in hadronic
collision, based on perturbative QCD (pQCD) calculations, seems to be
evasive \cite{p1}. This letter is an alternative attempt. We shows
that the color dipole approach can successfully describe inclusive photon
production in hadron-hadron collisions.

\section{Color dipole formalism}

The color dipole formalism, developed in \cite{bb} for the case of
the total and diffractive cross sections, can be also applied to
radiation \cite{hir}. Although in the process of electromagnetic
bremsstrahlung by a quark no dipole participates, the cross section
can be expressed via the more elementary cross section
$\sigma_{q\bar{q}}$ of interaction of a $\bar qq$ dipole.
Nevertheless, this is a fake, or effective dipole. Similar to a real
dipole, where color screening is provided by interactions with
either the quark or the antiquark, in the case of radiation the two
amplitudes for radiation prior or after the interaction screen each
other, leading to cancellation of the infra-red divergences
\cite{hir}.

The transverse momentum $p_{T}$ distribution of photon
bremsstrahlung in quark-nucleon interactions, integrated over the
final quark transverse momentum, was derived in \cite{p6} in terms
of the dipole formalism,
 \begin{eqnarray}
\frac{d \sigma^{qN}(q\to q\gamma)}{d(ln \alpha)d^{2}\vec{p}_{T}}&=&\frac{1}{(2\pi)^{2}}
\sum_{in,f}\sum_{L,T}
\int d^{2}\vec{r}_{1}d^{2}\vec{r}_{2}e^{i \vec{p}_{T}.(\vec{r}_{1}-\vec{r}_{2})}\nonumber\\
&\times&\phi^{\star T,L}_{\gamma q}(\alpha, \vec{r}_{1})
\phi^{T,L}_{\gamma q}(\alpha, \vec{r}_{2})
\Sigma_{\gamma}(x,\vec{r}_{1},\vec{r}_{2},\alpha),\nonumber\\  \label{m1}
 \end{eqnarray}
 where
 \begin{eqnarray}
\Sigma_{\gamma}(x,\vec{r}_{1},\vec{r}_{2},\alpha)&=&\frac{1}{2}\{
\sigma_{q\bar{q}}(x,\alpha r_{1})+\sigma_{q\bar{q}}(x,\alpha r_{2})\}\nonumber\\
&-&\frac{1}{2}\sigma_{q\bar{q}}(x,\alpha(\vec{r}_{1}-\vec{r}_{2})).\label{di}
\end{eqnarray}
%The appearance of the dipole cross section $\sigma_{q\bar{q}}(x,
%\alpha r)$ in the above equation is due to the quark displacement in
%the impact parameter plane after radiation of the photon.
 and $\vec{r}_{1}$ and $\vec{r}_{2}$ are the quark-photon transverse
 separations in the two radiation amplitudes contributing to the cross
 section, Eq.~(\ref{m1}), which correspondingly contains
 double-Fourier transformations. The parameter $\alpha$ is the
 relative fraction of the quark momentum carried by the photon, and is
 the same in both amplitudes, since the interaction does not change
 the sharing of longitudinal momentum. The transverse displacement
 between the initial and final quarks is $\alpha r_{1}$ and $\alpha
 r_{2}$ respectively. Since the amplitude of quark interaction has a
 phase factor $\exp(i\vec b\cdot\vec p_T)$, where $\vec b$ is the
 impact parameter of collision, the transverse displacement between
 the initial and final quarks leads to the color screening factor
 $1-\exp(i\alpha \vec{r}\cdot\vec{p}_{T})$.  In Eq.~(\ref{m1}) $T$
 stands for transverse and $L$ for longitudinal photons. The energy
 dependence of the dipole cross section, which comes via the variable
 $x=2 p_1
\cdot q/s$, where $p_1$ is the projectile four-momentum and $ q$ is
the four-momentum of the dilepton, is generated by additional
radiation of gluons which can be resummed in the leading $\ln(1/x)$
approximation.

In Eq.~(\ref{m1}) the light-cone (LC) wavefunction of the projectile
quark $\gamma q$ fluctuation has been decomposed into  transverse
$\phi^{T}_{\gamma q}(\alpha, \vec{r})$ and longitudinal
$\phi^{L}_{\gamma q}(\alpha, \vec{r})$ components,  and an average
over the initial quark polarization and sum over all final
polarization states of quark and photon is performed. These LC
wavefunction components $\phi^{T,L}_{\gamma q}(\alpha, \vec{r})$ can
be represented at the lowest order as:
\begin{eqnarray}
&&\sum_{in,f}\phi^{T\star}_{\gamma q}(\alpha, \vec{r}_{1})\phi^{T}_{\gamma q}(\alpha, \vec{r}_{2})
= \frac{\alpha_{em}}{2\pi^{2}}m^2_{q}\alpha^{4}K_{0}(\epsilon r_{1})K_{0}(\epsilon r_{2})\nonumber\\
&&+\frac{\alpha_{em}}{2\pi^{2}}[1+(1-\alpha)^{2}]\epsilon^{2}\frac{\vec{r}_{1}.\vec{r}_{2}}{r_{1}r_{2}}
K_{1}(\epsilon r_{1})K_{1}(\epsilon r_{2}),\nonumber\\
&&\sum_{in,f}\phi^{L\star}_{\gamma q}(\alpha, \vec{r}_{1})\phi^{L}_{\gamma q}(\alpha, \vec{r}_{2})=
\frac{\alpha_{em}}{\pi^{2}}M^{2}(1-\alpha)^{2}\nonumber\\
&&~~~~~~~~~~~~~~~~~~~~~~~~~~~~~~\times K_{0}(\epsilon r_{1})K_{0}(\epsilon r_{2}),\
\label{wave}
\end{eqnarray}
in terms of transverse separation $\vec{r}$ between photon $\gamma $
and quark $ q$ and the relative fraction $\alpha$ of the quark
momentum carried by the photon.  Here $K_{0,1}(x)$ denotes the
modified Bessel function of the second kind. We have also introduced
the auxiliary variable $\epsilon^{2}=\alpha^{2}
m_{q}^{2}+(1-\alpha)M^{2}$, where $M$ denotes the mass of dilepton
and $m_{q}$ is an effective quark mass which can be conceived as a
cutoff regularization.   This quark mass  has less influence on
dilepton production in $pp$ collisions, albeit it will be a
numerically important parameter for direct photon production, when
$M=0$. In general the quark mass $m_{q}$ should not be considered an
extra parameter. Indeed, depending on the kinematical variable $M$,
the Feynman variable $x_{F}$ and the square of the center of mass
energy of the colliding hadrons $s$, there always exists a range of
values of $m_{q}$ where the result does not depend on the specific
$m_q$ value. For direct photon $M=0$, $m_{q}$ cannot be zero since
the wave function becomes divergent. In this paper, as in Refs.
\cite{p6,mq}, we take $m_{q}=0.2$ GeV for both dilepton and direct
photon production. Notice also that $m_{q}$ is a more important
parameter in proton-nucleus collisions where a value of $m_{q}=0.2 $
GeV is needed in order to describe the nuclear shadowing effect
\cite{mqq}.

In order to obtain the hadron cross section from the elementary
partonic cross section Eq.~(\ref{m1}), one should sum up the
contributions from quarks and antiquarks weighted with the
corresponding parton distribution functions (PDFs) \cite{p6,mq},
\begin{eqnarray}
&&\frac{d \sigma^{DY}(pp\to \gamma^{\star} X)}{dM^{2}dx_{F}d^{2}\vec{p}_{T}}=
\frac{\alpha_{em}}{3\pi M^{2}}\frac{x_{1}}{x_{1}+x_{2}}\int_{x_{1}}^{1}\frac{d\alpha}{\alpha^{2}}\nonumber\\
&+&\sum Z_{f}^{2}\{q_{f}(\frac{x_{1}}{\alpha})+\bar{q}_{f}(\frac{x_{1}}{\alpha})\}
\frac{d \sigma^{qN}(q\to q\gamma^{\star})}{d(ln \alpha)d^{2}\vec{p}_{T}}\nonumber\\
&=&\frac{\alpha_{em}}{3\pi M^{2}(x_{1}+x_{2})}\int_{x_{1}}^{1}\frac{d\alpha}{\alpha}
F_{2}^{p}(\frac{x_{1}}{\alpha},Q)\frac{d \sigma^{qN}(q\to q\gamma^{\star})}{d(ln \alpha)d^{2}
\vec{p}_{T}}.\nonumber\\ \label{con0}
\end{eqnarray}
\begin{eqnarray}
&&\frac{d \sigma^{\gamma}(pp\to \gamma X)}{dx_{F}d^{2}\vec{p}_{T}}=
\frac{1}{x_{1}+x_{2}}\int_{x_{1}}^{1}\frac{d\alpha}{\alpha} F_{2}^{p}(\frac{x_{1}}{\alpha},Q)\nonumber\\
&&\hspace{2.5cm}\times\frac{d \sigma^{qN}(q\to q\gamma)}{d(ln \alpha)d^{2}\vec{p}_{T}}.\
\label{con}
\end{eqnarray}
The PDFs of the projectile enter in a combination which can be
written in terms of proton structure function $F_{2}^{p}$. Notice
that with our definitions the fractional quark charge $Z_{f}$ is not
included in the LC wave function of Eq.~(\ref{wave}), and that the
factor $\frac{\alpha_{em}}{3\pi M^{2}}$ in Eq.~(\ref{con0}) accounts
for the decay of the photon into the lepton pair. We use the
standard notation for the kinematical variables, $
x_{1}=(\sqrt{x_{F}^{2}+4\tau}+x_{F})/2$ denotes the momentum
fraction that the photon carries away from the projectile hadron in
the target frame, we define $x_{2}=x_{1}-x_{F}$,
$x_{F}=2p_{L}/\sqrt{s}$ is the Feynman variable and
$\tau=\frac{M^{2}+p^{2}_{T}}{s}$, where $p_{L}$ and $p_T$ denote the
longitudinal and transverse momentum components of the photon in the
hadron-hadron center of mass frame, $s$ is the center of mass energy
squared of the colliding protons and $M$ is the dilepton mass.
%The rapidity is defined as
%$y=\frac{1}{2}\ln\frac{\sqrt{x_{F}^{2}+4\tau}+x_{F}}{\sqrt{x_{F}^{2}+4\tau}-x_{F}}$. At
%midrapidity, we have $x_{F}=0$ and energy of photon
%$E_{\gamma}=p_{T}$. For midrapidity $y=0$, using the following
%relations $dx_{F}=2\tau dy$ and $dx_{F}=\frac{2}{\sqrt{s}}dp_{L}$ we obtain
%\begin{eqnarray}
%E_{\gamma}\frac{d^{3} \sigma}{d^{3}\vec{p}}&=&\int_{\frac{p_{T}}{\sqrt{s}}}^{1}\frac{d\alpha}
%{\alpha} F_{2}^{p}(\frac{p_{T}}{\alpha\sqrt{s}})\frac{d \sigma^{qN}(q\to q\gamma)}{d(ln \alpha)d^{2}\vec{p}_{T}}
%\nonumber\\
%\frac{d^{2} \sigma}{dy dp_{T}}&=&2\pi p_{T} E_{\gamma}\frac{d^{3} \sigma}{d^{3}\vec{p}}\
%\end{eqnarray}
We also need to identify the scale $Q$ entering in the proton
structure function in Eq.~({\ref{con}}), and relate the energy scale
$x$ of the dipole cross section entered in Eq.~(\ref{di}) to
measurable variables. From our previous definition, and following
previous works \cite{m2,mq} we have that $x=x_{2}$. At zero
transverse momentum, the dominant term in the LC wavefunction
Eq.~(\ref{wave}) is the one that contains the modified Bessel
function $K_{1}(\epsilon r)$. This function decay exponentially at
large values of the argument, so that the mean distances which
numerically contribute are of order $1/\epsilon$. On the other hand,
the minimal value of $\alpha$ is $x_{1}$, and therefore the
virtuality $Q^2$ which enters into the problem at zero transverse
momentum is $\sim (1-x_{1})M^{2}$. Thus the hard scale at which the
projectile parton distribution is probed turns out to be
$Q^{2}=p^{2}_{T}+ (1-x_{1})M^{2}$. Notice that in the previous
studies, $M^{2}$ \cite{mq} and $(1-x_{1})M^{2}$ \cite{m2} were used
for the scale $Q^{2}$. Nevertheless, these different choices for
$Q^{2}$ bring less than about a $20\%$ effect at small $x_{2}$
values.

The dipole cross section is theoretically unknown, although several
parametrizations have been proposed in the literature. For our
purposes, here we consider two parametrizations, the saturation
model of Golec-Biernat and W\"usthoff (GBW) \cite{gbw} and the
modified GBW coupled to DGLAP evolution (GBW-DGLAP) \cite{gbw-d}.

%\begin{figure}[!tbh]
%       \centerline{\includegraphics[width=8 cm] {figd.eps}}
%       \caption{Top: The dipole cross section for $x=10^{-2}, 10^{-4},
%       10^{-6}$ from right to left, solid lines correspond to the DGLAP
%       improved model and dashed lines the GBW saturation model. Down:
%       The gluon density as a function of $x$ for various dipole
%       size. The dipole size determines the evolution scale within the
%       gluon density. \label{fig-1}}
%\end{figure}

\subsection{GBW model parametrization}
In the GBW model \cite{gbw} the dipole cross section is parametrized as,
\begin{equation}
\sigma_{q\bar{q}}(x,\vec{r})=\sigma_{0}\left(1-e^{-r^{2}/R_{0}^{2}}\right), \label{gbw}
\end{equation}
where the parameters, fitted to DIS HERA data at small $x$, are
given by $\sigma_{0}=23.03$ mb, $R_{0}=0.4 \text{fm }\times
(x/x_{0})^{0.144}$, where $x_{0}=3.04\times 10^{-4}$. This
parametrization gives a quite good description of DIS data at
$x<0.01$.
%The saturation radius $R_{0}$ is analogous
%to the gluon distribution and determines the growth of cross section
%with decreasing $x$.
A salient feature of the model is that for decreasing $x$, the
dipole cross section saturates for smaller dipole sizes, and that at
small $r$, as perturbative QCD implies, $\sigma\sim r^{2}$ vanishes.
This is the so-called color transparency phenomenon \cite{bb}.
%For large dipole transverse
%size, dipole cross section saturates independent of energy and
%exponent in $R_{0}$.
One of the obvious shortcoming of  the GBW model is
that it does not match with QCD evolution (DGLAP) at large values of
$Q^{2}$. This failure can be clearly seen in the energy dependence
of $\sigma^{\gamma^{\star} p}_{tot}$ for $Q^{2}> 20~\text{GeV}^{2}$, where the
the model predictions are below the data \cite{gbw,gbw-d}.

\subsection{GBW couple to DGLAP equation and dipole evolution}
A modification of the GWB dipole parametrization model,
Eq.~(\ref{gbw}), was proposed in Ref.~\cite{gbw-d}
\begin{equation}
\sigma_{q\bar{q}}(x,\vec{r})=\sigma_{0}\left(1-exp\left(-\frac{\pi^{2}r^{2}
\alpha_{s}(\mu^{2})xg(x,\mu^{2})}{3\sigma_{0}}\right)\right), \label{gbw1}
\end{equation}
where the scale $\mu^{2}$ is related to the dipole size by
\begin{equation}
\mu^{2}=\frac{C}{r^{2}}+\mu_{0}^{2}. \label{scale}
\end{equation}
Here the gluon density $g(x,\mu^{2})$ is evolved to the scale
$\mu^{2}$ with the leading order (LO) DGLAP equation \cite{qcdnum}.
Moreover, the quark contribution to the gluon density is neglected
in the small $x$ limit, and therefore
\begin{equation}
\frac{\partial xg(x,\mu^{2})}{\partial \ln\mu^{2}}=\frac{\alpha_{s}(\mu^{2})}
{2\pi^{2}}\int_{x}^{1} dz P_{gg}(z)\frac{x}{z}
g(\frac{x}{z},\mu^{2}).
\end{equation}
where $P_{gg}(z)$ and $\alpha_{s}(\mu^{2})$ denote the QCD splitting
function and coupling, respectively.  The initial gluon density is
taken at the scale $Q_{0}^{2}=1 \text{GeV}^{2}$ in the form
\begin{equation} xg(x,\mu^{2})=A_{g}x^{-\lambda_{g}}(1-x)^{5.6},
\end{equation}
where the parameters $C=0.26$, $ \mu_{0}^{2}=0.52 \text{GeV}^{2}$,
$A_{g}=1.20$ and $\lambda_{g}=0.28$ are fixed from a fit to the DIS
data for $x<0.01$ and in a range of $Q^2$ between $0.1$ and $500$
$\text{GeV}^2$ \cite{gbw-d}.  We use the LO formula for the running
of the strong coupling $\alpha_{s}$, with three flavors and for
$\Lambda_{\text{QCD}}=0.2~\text{GeV}$. The dipole size determines
the evolution scale $\mu^{2}$ through Eq.~(\ref{scale}). The
evolution of the gluon density is performed numerically for every
dipole size $r$ during the integration of Eq.~(\ref{m1}). Therefore,
the DGLAP equation is now coupled to our master equations
(\ref{con0},\ref{con}). It is important to stress that the GBW-DGLAP
model preserves the successes of the GBW model at low $Q^{2}$ and
its saturation property for large dipole sizes, while incorporating
the evolution of the gluon density by modifying the small-$r$
behaviour of the dipole size.

The proton structure function in Eqs.~(\ref{con0},\ref{con}) is
parametrized as
\begin{equation}
F^{p}_{2}(x,Q)=A(x)\big[\frac{\ln(Q^{2}/\Lambda^{2})}{\ln(Q_{0}^{2}/\Lambda^{2})}\big]^{B(x)}
\left(1+\frac{C(x)}{Q^{2}}\right),
\end{equation}
with $Q_{0}^{2}=20 \text{GeV}^{2}$,  $\Lambda=0.25$ GeV, and the
functions $A(x), B(x)$ and $C(x)$ are parametrized in terms of $17$
parameters fitted to different experiments, and whose functional
forms can be found in the Appendix of Ref.~\cite{ps}. This parametrization is
only valid in the kinematic range of the data sets which cover
correlated regions in the ranges  $3.5\times 10^{-5}<x<0.85 $ and $0.2<Q^2<5000\text{GeV}^2$.

\section{numerical results}
Before we proceed to present the results in the color dipole
approach, some words regarding the validity of this formulation are
in order. Although both valence and sea quarks in the projectile are
taken into account through the proton structure function
Eqs.~(\ref{con0}, \ref{con}), the color dipole picture accounts only
for Pomeron exchange from the target, while ignoring its valence
content.  In terms of Regge phenomenology, this means that Reggeons
are not taken into account, and as a consequence, the dipole
approach predicts the same cross sections for both particle and
antiparticle induced DY reactions. Therefore, in principle this
approach is well suited for high-energy processes, i.e. small
$x_{2}$. The exact range of validity of the dipole approach is of
course not known a priori, but there is evidence \cite{mq,m2} in its
favor for values of $x_{2}<0.1$. In our case, however, we use a
parametrization of the dipole cross section fitted to DIS data for
Bjorken-$x<0.01$ and for energy scales $Q^2<500$. Given these
restrictions, at present there are not many data for DY cross
section at low $x_{2}$. Notice also that some data are integrated
over $x_{F}$ and $M$, and are therefore contaminated by
contributions not included into the color dipole approach.
\begin{figure}[!tbh]
       \centerline{\includegraphics[width=8 cm] {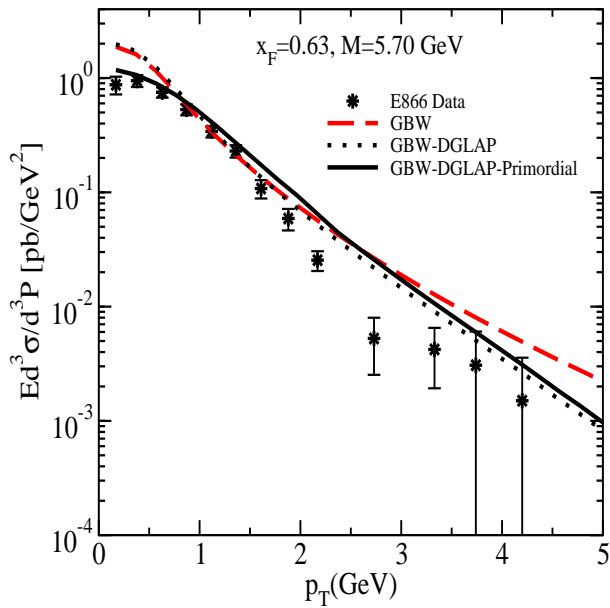}}
       \caption{The Dilepton spectrum in $800$-GeV $pp$ collisions at
       $x_{F}=0.63 $ and $M=5.7 $GeV. We show the result of the GBW
       dipole model (dashed line) and the GBW-DGLAP model (dotted
       line). We also show the result when a constant primordial
       momentum $\langle k_{0}^{2}\rangle=0.4 \text{GeV}^2$ is
       incorporated within the GBW-DGLAP dipole model (solid
       line). Experimental data are from Ref.~\cite{p2}. The E866
       error bars are the linear sum of the statistical and systematic
       uncertainties. An additional $\pm 6.5\%$ uncertainty in the
       experimental data points due to the normalization is also
       common to all points. \label{fig-2}}
\end{figure}
\begin{figure}[!tbh]
       \centerline{\includegraphics[width=8 cm] {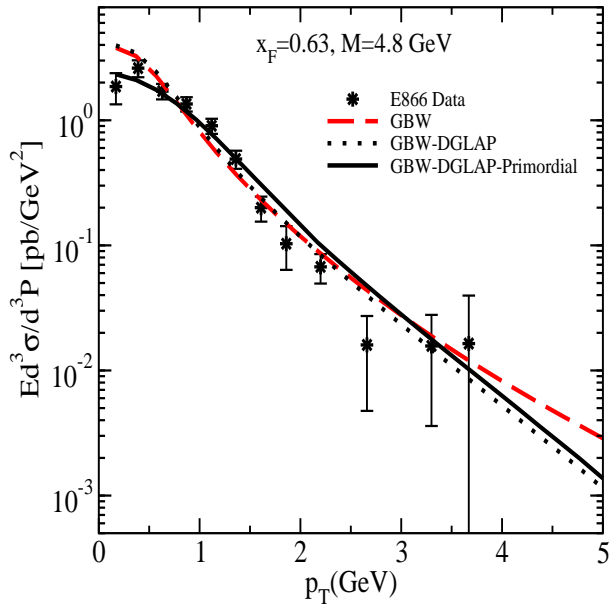}} \caption{
       The same as Fig.~\ref{fig-2}, except for $M=4.8$.\label{fig-3}}
\end{figure}

We compare the present approach to data for $800$-GeV $pp$
collisions from E886 \cite{p2}, which are not integrated over
$x_{F}$ and $M$, and correspond to the lowest $x_{2}$ values, i.e.
lightest $M$ and highest $x_{F}$. We selected a $x_{F}$ bin where
$0.55 < x_{F} < 0.8$, with an average value $\langle
x_{F}\rangle=0.63$. Within this bin we selected two bins with the
lightest average values for $M$, one for
$4.20<M_{\mu^{+}\mu^{-}}<5.20$, with an average value of $\langle
M_{\mu^{+}\mu^{-}}\rangle=4.80$ GeV, and the other for
$5.20<M_{\mu^{+}\mu^{-}}<6.20$, with an average value $\langle
M_{\mu^{+}\mu^{-}}\rangle=5.70$ GeV. The experimental data are
plotted, with errors, in Figs.~\ref{fig-2} and \ref{fig-3}. In our
calculations we have taken the experimental average values for
$x_{F}$ and $M$.

In Figs.~\ref{fig-2}, \ref{fig-3} we show the result obtained by the
dipole approach, for both the GBW and the GBW-DGLAP dipole models.
At low transverse momentum $p_{T}<2$ GeV both model predictions are
almost identical, but at higher $p_{T}$ the dipole parametrization
improved by  DGLAP evolution bends down towards the experimental
points improving the result. This is more obvious for higher values
of $M$. Notice that for the case of a smaller value of $x_{2}$ with
a lighter $M$, Fig.~\ref{fig-3}, where the dipole approach is better
suited, the GBW model without inclusion of the DGLAP evolution
already provides a good description of the data. We stress that the
theoretical curves in Figs.~\ref{fig-2}, \ref{fig-3} are the results
of a parameter free calculation. As we already pointed out, varying
the quark mass $m_{q}$ leaves the numerical results almost
unaffected. Notice also that in contrast to the LO parton model, no
$K$-factor was introduced, since the dipole parametrization fitted
to DIS data already includes contributions from higher order
perturbative corrections as well as non-perturbative effects
contained in DIS data.

One of the data point which surprisingly is left out from our
theoretical computation curves for both values of $M$, is the one at
the lowest $p_{T}$. In the dipole approach the DY cross section is
finite at $p_{T}=0$ due to the saturation of the dipole cross section,
which is in striking contrast to the LO pQCD correction to the parton
model, where one needs to resume the large logarithms
$\Ln(p_{T}^{2}/M^{2})$ from soft gluon radiation in order to obtain a
physically sensible results at $p_{T}=0$ \cite{pir}. One of the possible reasons
behind the lack of agreement between our result and the experimental
data at $p\to 0$ may be due to a soft non-perturbative primordial
transverse momentum distribution of the partons in the colliding
protons. Such a primordial transverse momentum may have various
non-perturbative origins, e. g. finite size effects of the hadron,
instanton effects, pion-cloud contributions. Moreover, in the parton
model it has been shown that even within the next-to-leading order
(NLO) pQCD correction, experimental data of heavy quark pair
production \cite{w1}, direct photon production \cite{w2} and DY lepton
pair production \cite{w3} can be only described if an average
primordial momentum as large as $1$ GeV is included (see also
Ref.~\cite{wang}). Such a large value for the initial transverse
momentum strongly indicates its perturbative origin in the parton
model, and in principle must have been already included in the pQCD
correction. Therefore, in the pQCD approach, it is still an open
question how to separate what is truly intrinsic and what is pQCD
generated transverse momentum. However, in the dipole approach all
perturbative and non-perturbative contributions, apart from the
finite-size effect of hadrons, are already encoded into the cross
section via fitting the dipole parameters to DIS data. Therefore, we
expect that in the dipole approach the primordial momentum should have
a purely non-perturbative origin, and to be considerably less than in
the parton model. One may introduce an intrinsic momentum contribution
in the following factorized form
\begin{equation}
\mathcal{F}(p_{T})\rightarrow \int d^{2}k_{T} \mathcal{F}(p_{T}-k_{T})\mathcal{G}_{N}(k_{T}),
\end{equation}
where the function $\mathcal{F}$ denotes the cross section defined
in Eqs.~(\ref{con0},\ref{con}). We assume that the initial $p_{T}$
distribution $\mathcal{G}_{N}(p_{T})$ has a Gaussian form,
\begin{equation}
\mathcal{G}_{N}(k_{T})=\frac{1}{\pi \langle k_{T}^{2}\rangle_{N}}e^{k_{T}^{2}/\langle k_{T}^{2}\rangle_{N}},
\end{equation}
where $\langle k_{T}^{2}\rangle_{N}$ is the square of the
two-dimensional width of the $p_{T}$-distribution for an incoming
quark, and also that $\langle k_{T}^{2}\rangle_{N}$ is a constant
independent of the hard scale $Q$, since the pQCD
radiation-generating transverse momenta are already taken into
account in our approach. The differential cross section convoluted
with the primordial momentum distribution in the GBW-DGLAP dipole
model are shown in Figs.~\ref{fig-2} and \ref{fig-3} with curves
denoted with GBW-DGLAP-Primordial.  A value around $\langle
k_{T}^{2}\rangle_{N}=0.4~ \text{GeV}^{2}$ can describe the
experimental points at low $p_{T}$ for both sets of data plotted in
Figs.~\ref{fig-2} and \ref{fig-3}. This value, as we expected, is
lower than the primordial momentum which has been used in the parton
model.

The experimental data points for $p_{T}\to 0$ should be taken with
some precaution, since there exists some disagreement between
different experiments for DY lepton pair production at low $p_{T}$.
Indeed, although the E772 and E866 measurements \cite{p2} have good
agreement among them over a wide range of values, they disagree at
$p_{T}\to 0$.
%The cross section at low $p_{T}$ is
%highly sensitive to the precise alignment of the beam.  This
%difference between two experiments might be due to the beam alignment
%issue in two experiments.
Therefore, the discrepancy between our theoretical results and
experimental data at $p_{T}\to 0$ might be just in fact an artifact
of the experiments.

Next we calculate the inclusive direct photon spectra within the
same framework. For direct photon we have $M=0$, and we assume again
a quark mass $m_{q}=0.2 $ GeV. As illustrative examples we compare
our results with the PHENIX and CDF experiments. Notice that direct
photon problem (with $M=0$), compared to the massive virtual photon
case (with $M$ as big as $\sim 5$ GeV), is numerically more involved
since the integrand in Eq.~(\ref{m1}) is divergent when $m_{q} \to
0$.
\begin{figure}[!tbh]
       \centerline{\includegraphics[width=8 cm] {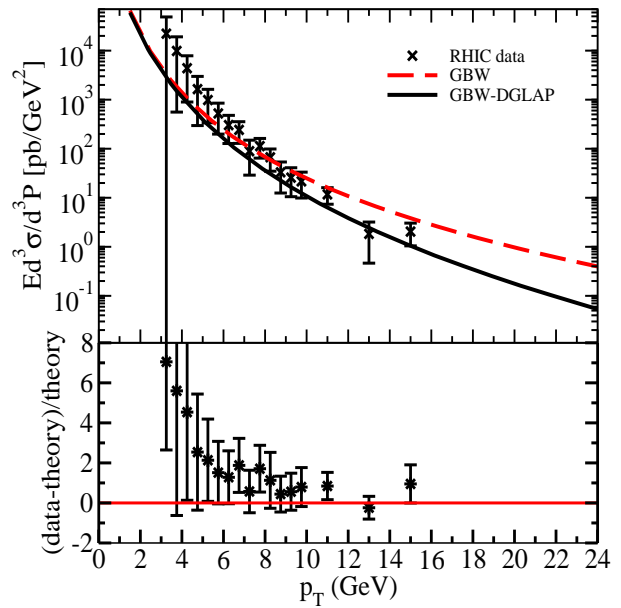}}
       \caption{Inclusive direct photon spectra obtained from the GBW
       and the GBW-DGLAP dipole models for midrapidity $\eta=0$ at
       RHIC energy $\sqrt{s}=200$ GeV.  Experimental data are from
       Ref.~\cite{p3}. In the down panel we use the GBW-DGLAP dipole
       model result for the theory. The error bars are the linear sum
       of the statistical and systematic uncertainties. \label{fig-4}}
\end{figure}
\begin{figure}[!tbh]
       \centerline{\includegraphics[width=8 cm] {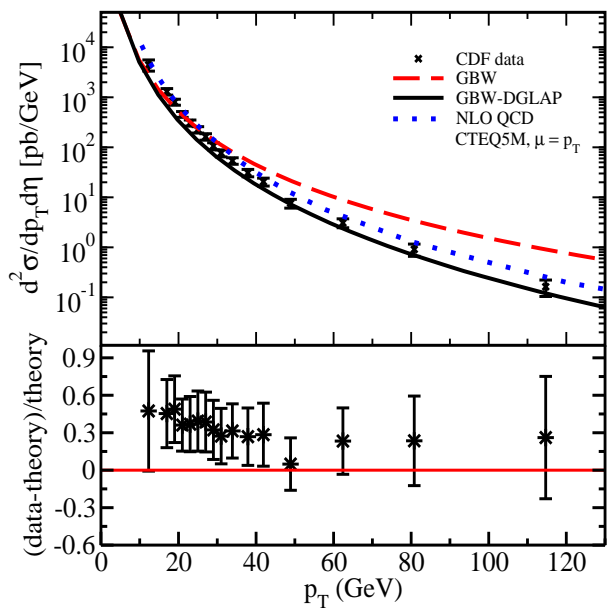}}
       \caption{ Inclusive direct photon spectra obtained from the GBW
       and GBW-DGLAP dipole models for midrapidity at CDF energy
       $\sqrt{s}=1.8$ TeV. Experimental data are for inclusive
       isolated photon from CDF experiment for $p\bar{p}$ collision at
       CDF energy and $|\eta|<0.9$ \cite{p4}. The NLO QCD curve is from the
       authors of reference \cite{ncdf} (given in table 3 of Ref.~\cite{newcdf}) and
       use the CTEQ5M parton distribution functions with the all
       scales set to the $p_{T}$. In the down panel we use 
       the GBW-DGLAP dipole model result for the theory. The error bars are
       the linear sum of the statistical and systematic
       uncertainties. \label{fig-5}}
\end{figure}

In Fig.~\ref{fig-4} we show the differential cross section obtained
from the GBW and the GBW-DGLAP dipole models at midrapidities, for
$pp$ collisions at RICH energies $\sqrt{s}=200 $ GeV. The experimental
data are from the PHENIX measurements for inclusive direct photon
production at $y=0$ \cite{p3}. We have also checked out that the
effect of the incorporation of the same transverse primordial momentum
$\langle k_{T}^{2}\rangle_{N}=0.4~\text{GeV}^{2}$ which can describe
the dilepton spectra at low $p_{T}$, will be in this case too small to
improve the results at the range of $p_{T}$ of the experimental
data. Without a physically sound guiding principle, however, the
introduction of a higher value of intrinsic momentum is somehow
unsatisfactory and will not be further discussed here.  Notice also
that, in contrast to the parton model, we have not included any photon
fragmentation function \cite{vv1,frag,vv2} for computing the cross
section, since the dipole formulation already incorporates all
perturbative (via Pomeron exchange) and non-perturbative radiation
contributions. It has been shown that the NLO pQCD prediction
\cite{vv1,vv2} are also consistent with the RHIC data within the
uncertainties \cite{p3}.

In Fig.~\ref{fig-5} we show the dipole models predictions for
inclusive prompt-photon production at midrapidities, and for CDF
energies $\sqrt{s}=1.8$ TeV. The experimental points are taken from
CDF data for inclusive isolated-photon, averaged over $|\eta|<0.9$
\cite{p4}. At lower transverse momentum  $p_{T}<30$ GeV the GBW
dipole model can reproduce rather fairly the experimental data, and at
higher $p_{T}$ values DGLAP evolution significantly improves the
results. %Notice, however
%that at such high $p_{T}$ neither parametrization of employed GBW
%dipole nor proton structure function is valid.
%Notice that at such a high $p_{T}$ even the GBW-DGLAP model
%parametrizations is not more valid.
In the collider experiments at the Tevatron, in order to reject the
overwhelming background of secondary photons which come from the
decays of pions, isolation cuts are imposed \cite{p4}. These cuts
affect the direct-photon cross section, in particular by reducing the
fragmentation effects.  Isolation conditions are not imposed in our
calculation, although the experimental data is for isolated
photon. However, it has been shown that the cross section does not
vary by more than $10\%$ under CDF isolation conditions and kinematics
\cite{cut}. Therefore, the main source of uncertainty in our approach
is due to the fact that the experimental points are averaged over
rapidity and contaminated by Reggeon contributions which are ignored
in the dipole approach.  One should also notice that the
parametrizations of the dipole cross section and proton structure
function employed in our computation have been fitted to data at
considerably lower $Q^{2}$ values (see previous section).
The NLO pQCD calculation for direct photon production at the Tevatron energy was
performed in Ref.~\cite{ncdf}.  
%The data show a steeper slope than the
%NLO pQCD calculation which cannot be explained by the systematic
%uncertainties of the measurement \cite{newcdf}. 
New independent measurement of direct photon at the Tevatron energy
which is in agreement with previous CDF measurement \cite{p4},
provided further evidence that the shape of the cross section as
function of $p_{T}$ cannot be fully described by the available NLO pQCD computation
\cite{newcdf}.

In this letter, we showed that both direct photon production and DY
dilepton pair production processes can be described within the same
color dipole approach without any free parameters. In contrast to the
parton model, in the dipole approach there is no ambiguity in defining
the intrinsic transverse momentum. Such a purely non-perturbative
primordial momentum improves the results in the case of dilepton pair
production, but does not play a significant role for direct photon
production at the given experimental range of $p_{T}$. We also showed
that the color dipole formulation coupled to the DGLAP evolution
provides a better description of data at large transverse momentum
compared to the GBW dipole model.

\section*{Acknowledgments}
The authors would like to thank T. Isobe for providing the experimental data in Ref.~\cite{p3}. 
This work was supported in part by Fondecyt (Chile) grants 1070517 and 1050519 and
by DFG (Germany)  grant PI182/3-1.

\end{document}